\documentclass[12pt]{article}

\usepackage{amssymb}

\usepackage{graphicx} 
\usepackage[hidelinks]{hyperref}
\usepackage{amsmath}
\usepackage{tikz}
\usetikzlibrary{arrows.meta}
\usetikzlibrary{shapes.geometric}
\usepackage{braket}
\usepackage{url}
\usepackage{authblk} 
\usepackage{bm}

\newcounter{bla}

\begin{document}

\title{\texttt{Simphony}: A full tight-binding package for lattice vibrations and topological phonon analysis}

\author[a,b,*]{Francesc Ballester}

\author[a,b,c]{Ion Errea}

\author[a,d,e]{Maia G. Vergniory}

\affil[a]{{Donostia International Physics Center, Donostia/San Sebastián 20018, Spain.}}
\affil[b]{{Department of Applied Physics, University of the Basque Country (UPV/EHU), Donostia/San Sebastián 20018, Spain.}}
\affil[c]{{Centro de Física de Materiales (CFM-MPC), CSIC-UPV/EHU, Donostia/San Sebastián 20018, Spain.}}
\affil[d]{{Département de Physique et Institut Quantique, Université de Sherbrooke, Sherbrooke, J1K 2R1 Québec, Canada}}
\affil[e]{{Regroupement Qu\'eb\'ecois sur les Mat\'eriaux de Pointe (RQMP), Quebec H3T 3J7, Canada}}
\affil[*]{Corresponding author: Francesc Ballester francisco.ballester@dipc.org}

\maketitle

\begin{abstract}

\texttt{Simphony} is an open-source software package designed for the topological analysis of lattice vibrations based on Wannier tight-binding models. Its primary function is to classify the topology of novel materials by computing bulk and slab phonon band structures, extracting phonon surface spectra, and providing analysis tools such as Wilson loop calculations and Weyl node detection. The workflow is analogous to that of established electronic topology codes like Wannier90 and WannierTools. It also incorporates long-range polar interactions during the wannierization process, making \texttt{Simphony} one of the first tools capable of diagnosing topology in polar insulators.

\end{abstract}

\section{Introduction}

Interest in topological materials has grown significantly in recent years, driven by the discovery of the quantum Hall effect~\cite{qhe,aqhe}, Weyl semimetals~\cite{Zhang-2020}, and exotic electronic surface states~\cite{reviewTI}. These phenomena are strongly protected by the symmetries of the system and remain robust in the presence of defects or impurities. Topologically protected surface states enable lossless current flow, which is essential for low-power electronic devices and promising for quantum computing applications.

Recently, the search for exotic states of matter has extended to the topological properties of lattice vibrations~\cite{topophodatabase, li-2021}, which could give rise to phonon-mediated topological effects such as superconductivity~\cite{Miceli-2022}  and dissipationless heat transport~\cite{Liu-2019} . Unlike electrons, phonons obey bosonic statistics, making the entire vibrational spectrum physically accessible, not just the states near the Fermi level. This has spurred interest in identifying symmetry-protected phonon modes~\cite{Zhang-2019, Komiyama-2022, Zhang-2022}, Weyl nodes in phonon band structures~\cite{zhang-2018, Zhang-2023}, and topological phonon surface states~\cite{martin-paper, wang-2021}, even at high energies within the phonon spectrum.

A key step in the search for novel topological phonon materials is the diagnosis of their topological character from bulk \textit{ab initio} calculations. This is commonly achieved using Topological Quantum Chemistry (TQC)~\cite{tqc}, which identifies topological phases based on the irreducible representations at high-symmetry points in the Brillouin zone. However, TQC may overlook band inversions or crossings along low-symmetry lines. To address these limitations, complementary methods based on Wannier functions are employed. These approaches enable the computation of Wilson loops and the construction of slab geometries to evaluate surface phonon dispersions and detect topological surface modes.

In electronic systems, a wide range of open-source software packages is available~\cite{Wannier90, WannierTools, pythtb} that can perform the wannierization process, compute Wilson loops, and construct slab Hamiltonians to calculate electronic surface states. In the case of lattice vibrations, some studies~\cite{zhang-2018} have leveraged the fact that phonons transform under the vector representation and therefore behave analogously to $p$ orbitals. This analogy enables the adaptation of electronic software to build tight-binding models that describe phononic systems. Such models correctly simulate phonons in bulk metallic materials and provide an alternative to standard Fourier interpolation for computing phonon spectra.

However, in non-metallic materials, unscreened charges give rise to macroscopic electric fields that generate long-range dipole–dipole interactions. These interactions induce an energetic splitting between longitudinal optical (LO) and transverse optical (TO) modes, known as LO–TO splitting. Because the dipole–dipole interaction is of infinite range, it cannot be treated via standard Fourier interpolation~\cite{xavierGonze1997}. Instead, this interaction is typically computed analytically, subtracted from the interatomic force constant (IFC) matrix before interpolation, and added back after the interpolation at arbitrary momentum points. 

This correction procedure is not implemented in the construction of tight-binding models for phonons using the aforementioned electronic software. Previous studies of phonon band topology~\cite{topophodatabase, li-2021} have often neglected the long-range dipole–dipole interaction in polar materials, arguing that the LO–TO splitting does not affect the irreducible representation content of the band representation. However, we argue that this interaction may still alter the topological nature of the phonon band structure by inducing band inversions, crossings, or reordering of band groupings—features that are not captured by symmetry labels alone.

We present \texttt{Simphony}, an open-source software package for the topological classification of lattice vibrations using tight-binding models—applicable to a broad range of systems, including those exhibiting LO–TO splitting. \texttt{Simphony} incorporates long-range dipole–dipole interactions in a manner consistent with state-of-the-art \textit{ab initio} codes such as \textsc{Quantum ESPRESSO} ~\cite{QE-2009,QE-2017}, VASP~\cite{VASP-1,VASP-2,VASP-3}, and \textsc{Abinit}~\cite{Abinit-1,Abinit-2}. The code can compute bulk phonon band structures, perform Wilson loop calculations, and construct slab Hamiltonians to diagnose topological properties. Additionally, it is capable of identifying nodal points in bulk spectra, calculating the monopole charge associated with Weyl nodes~\cite{Zhang-2023}, and generating surface energy contours to visualize topological phonon surface states.

The manuscript is organized as follows. In Section~\ref{theory}, we introduce the theoretical framework and methods implemented in \texttt{Simphony}. Section~\ref{usage} provides instructions for compiling the software and describes its usage. Finally, in Section~\ref{examples}, we demonstrate the capabilities of \texttt{Simphony} through three illustrative examples of topological phonon systems.

\section{Theory and Methods}
\label{theory}

In \textit{ab initio} methods, there are two primary approaches to compute the interatomic force constants (IFCs): in reciprocal space or in real space. In both cases, the calculations are performed on a coarse grid and require Fourier interpolation to obtain the phonon spectra on a finer grid in reciprocal space. While this interpolation is straightforward for metallic systems, it becomes more involved in polar materials due to the presence of unscreened long-range dipole–dipole interactions, which cannot be directly Fourier interpolated. Instead, these interactions must be analytically removed prior to interpolation and added back afterward~\cite{BaroniDFPT,xavierGonze1997}. The first part of this section is dedicated to explaining how \texttt{Simphony} constructs the Wannier tight-binding model and incorporates the long-range dipole–dipole interaction after interpolation.

In the second part, we focus on the tools available in \texttt{Simphony} for the topological classification of phonon bands. These include numerical techniques such as the computation of Wannier charge centers (via Wilson loops), the construction of slab Hamiltonians, and the analysis of surface energy contours using an iterative Green's function method. We detail how each of these methods is implemented from Wannier-based tight-binding models within \texttt{Simphony}.

\subsection{Dynamical Matrix, Fourier Interpolation and Long-Range Interactions}

Let us consider a supercell of a periodic crystal made of $N$ unit cells. We can denote the atomic positions as
\begin{equation}
    \bm{r}^a_\kappa = \bm{R}_a + \bm{\tau}_\kappa + \bm{u}^a_\kappa \text{,}
\end{equation}
where $\bm{R}_a$ is the lattice vector of the $a$-th unit cell, $\bm{\tau}_\kappa$ is the equilibrium position vector of the $\kappa$-th atom inside the unit cell, and $\bm{u}^a_\kappa$ is the displacement vector from the equilibrium position of the $\kappa$-th atom in the $a$-th unit cell. In the harmonic approximation, the Born–Oppenheimer energy $E_{BO}$ of the crystal is Taylor-expanded around the minimum with respect to the displacements. The dynamical matrix can then be obtained as
\begin{equation}
    D_{\kappa\alpha,\kappa'\beta}(a,b) = \frac{1}{\sqrt{M_\kappa M_{\kappa'}}} \frac{\partial^2 E_{BO}}{\partial u_{\kappa,\alpha}^a \partial u_{\kappa',\beta}^b} \Bigg|_{\bm{u}=\bm{0}}\text{,}
    \label{eq:dynmat}
\end{equation}
where $M_\kappa$ is the mass of the atom labeled with $\kappa$, and $\alpha$ and $\beta$ are Cartesian indexes. The dynamical matrix can be Fourier interpolated into any point in the reciprocal space by
\begin{equation}
    D_{\kappa\alpha,\kappa'\beta}(\bm{q}) = \frac{1}{N}\sum_{a,b}\exp{(-i\bm{q}\cdot(\bm{R}_a-\bm{R}_b))} D_{\kappa\alpha,\kappa'\beta}(a,b)  = \sum_{b}\exp{(i\bm{q}\cdot\bm{R}_b)}D_{\kappa\alpha,\kappa'\beta}(0,b)  \text{,}
    \label{eq:fourierdynmat}
\end{equation}
where we have used translational invariance for the last equation.

In practice, if the real-space dynamical matrix in Eq.~\eqref{eq:dynmat} is constructed in a supercell containing $N$ unit cells, Eq.~\eqref{eq:fourierdynmat} yields an exact result for all $N$ $\bm{q}$-points that are commensurate with the supercell. The dynamical matrix at any other $\bm{q}$-point must be obtained via Fourier interpolation. Typically, dynamical matrices are computed on a regular $q_1 \times q_2 \times q_3$ grid in the Brillouin zone by generating a commensurate $q_1 \times q_2 \times q_3$ supercell and calculating the interatomic force constants (IFCs).

In insulating materials, the absence of screening and the localization of charge around ionic positions lead to a nonzero polarization vector when atoms are displaced. This results in macroscopic electric fields that contribute to the total energy of the system through long-range dipole–dipole interactions. These electric fields particularly affect the longitudinal optical (LO) modes, increasing their energy relative to the transverse optical (TO) modes.

For lattice vibrations, the dipole–dipole interaction can be analytically incorporated into the dynamical matrix~\cite{xavierGonze1997}. The analytical expression depends on two key parameters of the system: the electronic dielectric permittivity tensor and the Born effective charge tensor of each atom. The former relates the macroscopic displacement field to the macroscopic electric field at low frequencies:
\begin{equation}
    \varepsilon^\infty_{\alpha\beta} = \frac{\partial\mathcal{D}_{\text{mac},\alpha}}{\partial\mathcal{E}_{\text{mac},\beta}}\text{.}
\end{equation}
The latter is defined as the proportionality coefficient between the polarization per unit cell along direction $\beta$ and the displacement of atom $\kappa$ along Cartesian direction $\alpha$, under zero electric field and to linear order:
\begin{equation}
    Z^*_{\kappa,\beta\alpha} = \Omega_0\frac{\partial\mathcal{P}_{\text{mac},\beta}}{\partial u_{\kappa,\alpha}} \Big|_{\bm{q} = \bm{0}}\text{,}
\end{equation}
where $\Omega_0$ is the volume of the unit cell. Both quantities can be computed  \textit{ab initio} using standard packages like Quantum ESPRESSO \cite{QE-2009, QE-2017} or VASP \cite{VASP-1,VASP-2,VASP-3}.

The long-range dipole-dipole interactions present nonanalytical behavior close to $\Gamma$, which translates into an approximate $1/d^3$ decay, where $\bm{d}^a_{\kappa,\kappa'}=\bm{R_a}+\bm{\tau}_{\kappa'}-\bm{\tau}_\kappa$. The long-range interactions do not vanish outside the supercell and, therefore, cannot be Fourier interpolated. This is fixed by removing the long-range interactions before interpolation and adding them back afterwards, which is the standard procedure in most codes \cite{phonopy-1, phonopy-2}.

In fact, the dynamical matrix (in real space) can be separated into two terms: a short-range (SR) and a long-range (LR) term:
\begin{equation}
    D_{\kappa\alpha,\kappa'\beta}(0,a) = D^{\text{SR}}_{\kappa\alpha,\kappa'\beta}(0,a) + D^{\text{LR}}_{\kappa\alpha,\kappa'\beta}(0,a)\text{.}
\end{equation}
The $D^{\text{SR}}_{\kappa\alpha,\kappa'\beta}(0,a)$ term is the one that must be Fourier interpolated, while the long-range (LR) term is added after the interpolation. The LR term in real space is given by~\cite{xavierGonze1997}:
\begin{equation}
D^{\text{LR}}_{\kappa\alpha,\kappa'\beta}(0,a) = \frac{1}{\sqrt{M_\kappa M_{\kappa'}}}  \frac{1}{(\det\hat{\varepsilon}^{\infty})}\sum_{\alpha' \beta'} Z^*_{\kappa,\alpha\alpha'}Z^*_{\kappa',\beta\beta'} \bigg(\frac{(\varepsilon^{\infty})^{-1}_{\alpha'\beta'}}{(\mathfrak{D}^a_{\kappa,\kappa'})^3}-3\frac{\Delta_{\alpha'}\Delta_{\beta'}}{(\mathfrak{D}^a_{\kappa,\kappa'})^5}\bigg)
    \label{eq:lrrealspace}
\end{equation}
with $\Delta^a_{\kappa,\kappa'\alpha} = \Sigma_{\beta}(\varepsilon^{\infty})^{-1}_{\alpha\beta}d^a_{\kappa,\kappa'\beta}$ and $\mathfrak{D}^a_{\kappa,\kappa'} = \sqrt{\bm{\Delta}^a_{\kappa,\kappa'}\cdot\bm{d}^a_{\kappa,\kappa'}}$, so the dielectric tensor works as a metric in real space. On the other hand, in reciprocal space, Eq. \eqref{eq:lrrealspace} can be written as:
\begin{multline}
   D^{\text{LR}}_{\kappa\alpha,\kappa'\beta}(\bm{q}) = \frac{1}{\sqrt{M_\kappa M_{\kappa'}}} \sum_{\bm{G} \text{with} \bm{K} = \bm{G}+\bm{q}}\frac{4\pi}{\Omega_0}\frac{(\sum_{\alpha'}K_{\alpha'}Z^*_{\kappa,\alpha'\alpha})(\sum_{\beta'}K_{\beta'}Z^*_{\kappa',\beta'\beta})}{\sum_{\gamma\gamma'} K_\gamma \varepsilon^\infty_{\gamma\gamma'}K_{\gamma'}} \exp{(i\bm{K}\cdot(\bm{\tau}_\kappa - \bm{\tau}_{\kappa'}))} \\
    -\delta_{\kappa\kappa'}\frac{1}{\sqrt{M_\kappa M_{\kappa'}}}\sum_{\kappa''}\sum_{\bm{G} \neq \bm{0}}\frac{4\pi}{\Omega_0}\frac{(\sum_{\alpha'}G_{\alpha'}Z^*_{\kappa,\alpha'\alpha})(\sum_{\beta'}G_{\beta'}Z^*_{\kappa'',\beta'\beta})}{\sum_{\gamma\gamma'} G_\gamma \varepsilon^\infty_{\gamma\gamma'}G_{\gamma'}} \exp{(i\bm{G}\cdot(\bm{\tau}_\kappa - \bm{\tau}_{\kappa''}))}\text{.}
    \label{eq:lrdynqspace}
\end{multline} 
In practice, this term is computed using an Ewald Sum technique \cite{xavierGonze1997}.

The nonanalytic behavior at $\Gamma$ of the LR interaction is clear from the previous equation. The limit at $\Gamma$ can be computed analytically as:
\begin{equation}
    D^{\text{LR}}_{\kappa\alpha,\kappa'\beta}(\bm{q} \rightarrow \bm{0}) = \frac{1}{\sqrt{M_\kappa M_{\kappa'}}}\frac{4\pi}{\Omega_0} \frac{(\bm{q}\cdot \hat{Z}^*_{\kappa})_\alpha(\bm{q}\cdot \hat{Z}^*_{\kappa'})_\beta}{\bm{q}\cdot\hat{\varepsilon}^{\infty}\cdot\bm{q}}\text{,}
\end{equation}
which is added to the full dynamical matrix at $\Gamma$ after interpolation. This limit is direction-dependent, and the eigenvalues along different directions do not necessarily coincide in non-cubic systems.

The dynamical matrix is then Fourier interpolated as in Eq.\eqref{eq:fourierdynmat}, considering only the short-range (SR) term, while the long-range (LR) term is added back afterward at any $\bm{q}$ point by explicitly using Eq.\eqref{eq:lrdynqspace}.

\subsection{Wannier Tight-Binding for Phonons}

The above-mentioned Fourier interpolation process can be reformulated within a tight-binding (TB) framework, which is less commonly applied to phonon systems than to electronic ones. A TB model is constructed by projecting the system's Hamiltonian onto a local basis of orbitals. For phonons, the role of the Hamiltonian is played by the dynamical matrix, which naturally translates the Fourier interpolation procedure into the TB formalism.
The key challenge lies in identifying an appropriate basis onto which the dynamical matrix can be projected in order to construct the TB Hamiltonian. The choice of basis determines the nature of the resulting TB model. In this work, we adopt a basis of Maximally Localized Wannier Functions (MLWFs), which, in the context of lattice vibrations, are referred to as Maximally Localized Lattice Wannier Functions (MLLWFs)~\cite{reviewWannier}.

The $\mu$-th phonon eigenmode and eigenfrequency at $\bm{q}$ can be obtained using the eigenvalue equation
\begin{equation}
    \bm{D}(\bm{q}) \bm{e}_{\mu,\bm{q}} = \omega^2_{\mu,\bm{q}}  \bm{e}_{\mu,\bm{q}}\text{.}
\end{equation}
Phonon eigenmodes are, by definition, delocalized. However, it is possible to construct a localized basis from the delocalized set of normal modes~\cite{Kohn1973, RabeWaghmare1995}, resulting in a set of Lattice Wannier Functions (LWFs). In the electronic case, this process is known as Wannierization and is typically performed using \textsc{Wannier90}\cite{Wannier90}, which generates a basis of Maximally Localized Wannier Functions (MLWFs) $w^a_m(\bm{r})$ from a set of Bloch functions $\psi_{\mu,\bm{k}}$ corresponding to an isolated group of bands\cite{reviewWannier}, via a generalized Fourier transform that includes band mixing.
\begin{equation}
    w^a_\mu(\bm{r}) = \frac{1}{\sqrt{N}}\sum_{\nu\bm{k}} \exp{(-i\bm{k}\cdot\bm{R}_a)} U_{\mu\nu\bm{k}} \psi_{\nu,\bm{k}}\text{,}
\end{equation}
where the matrix $U_{\mu\nu\bm{k}}$ is responsible for the band mixing. If $U_{\mu\nu\bm{k}}$ is unitary, the resulting MLWFs are orthonormal. The problem of maximal localization is then reduced to finding a $U_{\mu\nu\bm{k}}$ that minimizes the spread of each $w^a_\mu(\bm{r})$. The inverse of this transformation can be expressed as
\begin{equation}
     \psi_{\mu,\bm{k}} = \frac{1}{\sqrt{N}} \sum_{\nu a} \exp{(i\bm{k}\cdot\bm{R}_a)} U^\dagger_{\mu\nu\bm{k}} w^a_\nu(\bm{r}) \text{.}
     \label{eq:inverseWF}
\end{equation}
For the case of MLLWFs, we consider a vibrational eigenmode, defined as $e^a_{\mu,\bm{q}\kappa\alpha} \equiv \frac{1}{\sqrt{N}}\exp{(i\bm{q}\cdot\bm{R}_a)} e_{\mu,\bm{q}\kappa\alpha}$, which can be written as
\begin{equation}
    e^a_{\mu,\bm{q}\kappa\alpha} = \frac{1}{\sqrt{N}} \sum_{b\kappa'\beta} \exp{(i\bm{q}\cdot\bm{R}_b)}\delta_{ab}\delta_{\kappa\kappa'}\delta_{\alpha\beta} e_{\mu,\bm{q}\kappa'\beta} \text{.}
\end{equation}
By noting the similarities between this equation and Eq. \eqref{eq:inverseWF} it can be deduced that the individual displacement of the atoms $\delta_{ab}\delta_{\kappa\kappa'}\delta_{\alpha\beta}$ correspond to the MLLWFs \cite{Giustino2007}, and since lattice vibrations transform under the vector representation, MLLWFs are analogues to a set of three $p$ orbitals per atom.

The projection of the dynamical matrix onto this set of MLLWFs  leads to a straightforward definition of the phonon TB Hamiltonian:
\begin{equation}
    H_{\kappa\alpha,\kappa'\beta}(\bm{R}_a) =  D_{\kappa\alpha,\kappa'\beta}(0,a)  \text{.}
\end{equation}
For metallic systems, the previous equation can be implemented without considering long-range interactions. In the case of polar materials, the long-range (LR) part of the dynamical matrix must be removed before any type of interpolation can be performed to obtain the real-space dynamical matrix. Therefore, in these cases, the tight-binding (TB) Hamiltonian must be constructed by projecting the short-range (SR) part of the dynamical matrix onto the basis of Maximally Localized Lattice Wannier Functions (MLLWFs), i.e.,
\begin{equation}
    H_{\kappa\alpha,\kappa'\beta}(\bm{R}_a) =  D^{\text{SR}}_{\kappa\alpha,\kappa'\beta}(0,a)  \text{.}
    \label{eq:hamdynmatSR}
\end{equation}
For clarity, we will drop the index $a$ from now on. Furthermore, when the indices $\kappa$ ($\kappa'$) and $\alpha$ ($\beta$) are not explicitly necessary, we will combine them into a single index $m$ ($n$).

In order to obtain the TB Hamiltonian at any arbitrary point in reciprocal space, a Fourier interpolation is performed. In the case of polar materials, the long-range (LR) part of the dynamical matrix must then be explicitly added back at the desired point:
\begin{equation}
    H_{mn}(\bm{k}) = D^{\text{LR}}_{mn}(\bm{k})+\sum_{\bm{R}} \exp{(i\bm{k}\cdot\bm{R})}H_{mn}(\bm{R}) \text{.}
    \label{eq:latticegauge}
\end{equation}

\subsection{Slabs}

In order to study the surface states of a system, it is useful to construct slab structures, which break periodic boundary conditions along one direction. This can be achieved either by specifying a surface using Miller indices or by defining two new surface vectors. Both approaches are equivalent to defining a rotation matrix $U$, which transforms the original set of lattice vectors ${\bm{a}_1, \bm{a}_2, \bm{a}_3}$ into a new set ${\bm{a}'_1, \bm{a}'_2, \bm{a}'_3}$.

The surface of the slab is defined by $\{\bm{a}'_1, \bm{a}'_2\}$. The third basis vector is obtained by
\begin{equation}
\label{eq:r3}
    \bm{a}'_3 = g^{\frac{2}{3}}\big( (A^T)^{-1} (\bm{a}'_1 \times \bm{a}'_2)\big)
\end{equation}
where $g=\sqrt{\det A}$ and $A$ is defined as
\begin{equation}
    A = \begin{pmatrix}
        \bm{a}_1\\ \bm{a}_2\\ \bm{a}_3
    \end{pmatrix} \begin{pmatrix}
        \bm{a}_1& \bm{a}_2& \bm{a}_3
    \end{pmatrix} \text{.}
\end{equation}
Which ensures that $\bm{a}'_3$ respects the geometry of the bulk lattice and maintains the unit cell volume \cite{ase}. In the case of cubic systems, Eq. \eqref{eq:r3} reduces to a vector product scaled by the volume of the unit cell.

We now consider $\bm{a}'_1$ and $\bm{a}'_2$ to be the periodic directions and repeat the rotated unit cell along $\bm{a}'_3$ $n_s$ times. This will lead to a slab Hamiltonian with block structure, where the blocks can be labeled by the $i,j$ indices, both running from $1$ to $n_s$. To obtain the block matrices, labeled as $H_{mn,ij}$, we consider the 2D momentum in the surface Brillouin zone $\bm{k}_{||}$. Initially, we have a value for each $H_{mn}$ on a grid of real space vectors $\bm{R}_{grid}$, written in the basis $\{\bm{a}_1, \bm{a}_2, \bm{a}_3\}$. We write the set of $\bm{R}_{grid}$ in the new basis $\{\bm{a}'_1, \bm{a}'_2, \bm{a}'_3\}$, where the different vectors can be written in direct coordinates as $\bm{R} = (n'_1,n'_2,n'_3)$. The block matrix $H_{mn,ij}$ corresponds to the Fourier transform performed in the subspace of $\bm{R}_{grid}$ defined by $n'_3 = j-i$
\begin{equation}
    H_{mn,ij}(\bm{k}_{||}) = \sum_{n'_1,n'_2} \exp{(i\bm{k}_{||}\cdot(n'_1\bm{a}'_1 + n'_2\bm{a}'_2))} H_{mn}(n'_1\bm{a}'_1 + n'_2\bm{a}'_2 + (j-i)\bm{a}'_3)\text{.}
    \label{eq:surfhamij}
\end{equation}
Given the challenges associated with interpolating the long-range part of the dynamical matrix, we use Eq.~\eqref{eq:lrdynqspace} to compute the dipole–dipole term in momentum space on a very dense grid of points. By subsequently performing a Fourier transform into real space from this dense grid, we suppress the numerical errors introduced during interpolation due to the nonanalytic behavior of the long-range dipole–dipole interaction. After calculating the long-range contribution for each vector in $\bm{R}_{\text{grid}}$, we include each term in the interpolation as:
\begin{equation}
\begin{aligned}
    H_{mn,ij}(\bm{k}_{||}) = \sum_{n'_1,n'_2} \exp{(i\bm{k}_{||}\cdot(n'_1\bm{a}'_1 + n'_2\bm{a}'_2))} \big(H_{mn}(n'_1\bm{a}'_1 + n'_2\bm{a}'_2 + (j-i)\bm{a}'_3) +\\
    D^{LR}_{mn}(n'_1\bm{a}'_1 + n'_2\bm{a}'_2 + (j-i)\bm{a}'_3)\big)\text{,}
\end{aligned}
    \label{eq:surfhamijloto}
\end{equation}
thus mitigating its nonanalytic behavior \cite{Sohier-2017,Sohier-2017-2}.

The whole Hamiltonian for a system with $n_s$ slabs stacked along the $\bm{a}'_3$ direction is
\begin{equation}
    H^{slab}_{mn}(\bm{k_{||}}) = 
    \begin{pmatrix}
        H_{mn,11}(\bm{k}_{||}) & H_{mn,12}(\bm{k}_{||}) & \dots & H_{mn,1n_s}(\bm{k}_{||})\\
        H_{mn,21}(\bm{k}_{||}) & H_{mn,22}(\bm{k}_{||}) & \dots & H_{mn,2n_s}(\bm{k}_{||})\\
        \vdots & \vdots &\ddots & \vdots\\
        H_{mn,n_{s}1}(\bm{k}_{||}) & H_{mn,n_{s}2}(\bm{k}_{||}) & \dots & H_{mn,n_{s}n_s}(\bm{k}_{||})
    \end{pmatrix}
\end{equation}
and we can get the surface states dispersion by diagonalizing this Hamiltonian. We can also project the eigenmodes $\bm{e}_{\mu,\bm{k_{||}}}$ of the Hamiltonian onto the surface atoms to assign a surface weight $p_{\mu,\bm{k_{||}}}$ for each mode, that is
\begin{equation}
    p_{\mu,\bm{k_{||}}} = \sum_{m\in\text{surface}} |e_{\mu,\bm{k_{||}},m}|^2 \text{.}
    \label{eq:projectionSS}
\end{equation}
In this way, we can easily identify the surface modes by their associated weight. Since we consider normalized eigenmodes, the closer $p_{\mu,\bm{k_{||}}}$ is to one, the more localized the mode $\bm{e}{\mu,\bm{k{||}}}$ is at the surface.

\subsection{Wannier Charge Center}

The Wilson loop method~\cite{yu-2011} has been widely used to diagnose the topology of materials~\cite{Thouless-2016,Kane-2005,Fu-2007}. In this work, we employ an equivalent approach: the Wannier charge center (WCC) method~\cite{soluyanov-2011, Gresch-2017}. This technique is based on tracking the evolution of the expectation value of the position operator within a unit cell centered at the origin. The WCC is closely related to the Berry connection and involves momentum-space derivatives of the bulk eigenmodes, making it sensitive to gauge transformations of the form $\bm{e}{\mu}(\bm{k}) \rightarrow e^{-i\beta(\bm{k})}\bm{e}{\mu}(\bm{k})$, where $\beta(\bm{k})$ is a continuous real function.

For WCC calculations, we adopt a gauge different from that in Eq.~\eqref{eq:latticegauge}, explicitly accounting for the atomic positions $\bm{\tau}_\kappa$.
\cite{pythtb}:
\begin{equation}
    H_{\kappa\alpha,\kappa'\beta}(\bm{k}) = \sum_{\bm{R}} \exp{(i\bm{k}\cdot(\bm{R}+\bm{\tau}_\kappa - \bm{\tau}_{\kappa'}))}H_{\kappa\alpha,\kappa'\beta}(\bm{R}) \text{.}
    \label{eq:atomicgauge}
\end{equation}
The eigenvalues are unchanged under this gauge transformation, but the eigenmodes $\bm{e}_{\mu,\bm{k}}$ are analogous to the periodic part of the Bloch wave function in the electronic case. The long-range interaction is added back after interpolation in the same way as in Eq. \eqref{eq:latticegauge}.

In three dimensions  WCCs are computed on a plane of the first Brillouin zone defined by the $\bm{k}_1$ and $\bm{k}_2$ directions and a center point $\bm{k}_0$, with both $\bm{k}_1$ and $\bm{k}_2$ unitary vectors. This defines a general point inside the plane $\bm{k} = q_1\bm{k}_1 + q_2\bm{k}_2 + \bm{k}_0$, which can simply be labeled as $(q_1,q_2)$. Then we compute the expectation value of the position operator along direction $\bm{k}_2$ as a function of $q_1$
\begin{equation}
    \bar{y}_\mu (q_1) = \frac{i}{2\pi} \int_{-\pi}^{\pi} dq_2\ \bm{e}^\dagger_{\mu,(q_1,q_2)} \partial_{q_2} \bm{e}_{\mu,(q_1,q_2)}
\end{equation}
with $\bar{y}_\mu (q_1)$ being modulo 1. The values of $\bar{y}_\mu (q_1)$ are not necessarily gauge invariant, but the sum over $q_1$ must be invariant \cite{King-1993}. 

To implement this definition, we follow the algorithm present in Refs. \cite{Gresch-2017, WannierTools} and developed in Ref.\cite{soluyanov-2011}. We discretize the grid along the direction of $\bm{k}_2$, where we label $q_2^{j}$ as the coefficient of the $j$-th point at the given grid. This way we can define a set of overlap matrices, labeled by the index $j$
\begin{equation}
    M^{j}_{\mu\nu} (q_1) = \bm{e}^\dagger_{\mu,(q_1,q_2^j)} \bm{e}^{\phantom{\dagger}}_{\nu,(q_1,q_2^{j+1})} \text{,}
\end{equation}
where each matrix is a function of $q_1$. $M^{j}_{\mu\nu} (q_1)$ does not not follow gauge invariance, so we impose it by first performing a single value decomposition on the overlap matrix $M=V\Sigma W^\dagger$, where $\Sigma$ is a positive real diagonal matrix and $V$ and $W$ are unitary matrices. By rotating the eigenmodes at $q_2^{j+1}$ by the unitary rotation matrix $WV^\dagger$, $M^{j}_{\mu\nu} (q_1)$ becomes gauge invariant under a summation over $q_1$. The modes at $q_2 = \pi$ and the modes at $q_2 = -\pi$ are related by a unitary rotation matrix
\begin{equation}
    \Lambda = \prod_j W^j(V^j)^\dagger
\end{equation}
whose eigenvalues are $\lambda_\mu = e^{-i\bar{y}_\mu}$.

Finally, we obtain the WCC by
\begin{equation}
    \bar{y}_\mu(q_1) = -\frac{1}{2\pi} \operatorname{Im}\ln \lambda_\mu\text{.}
\end{equation}

\subsection{Green's Function for Surface State Calculation}

To study symmetry-enforced nodal lines and surface states between nodal points, an energy contour is typically computed. By analyzing the spectral function of the surface Green's function, one can obtain the density of states (DOS) at a specified point and energy range in the Brillouin zone.

In \texttt{Simphony}, we use an iterative Green's function algorithm described in Ref.\cite{M-P-Lopez-Sancho_1985} and employed in other software\cite{WannierTools} for similar purposes in electronic systems. The primary objective of the algorithm is to determine an effective principal layer that is sufficiently large for interactions between neighboring principal layers to become negligible. The Green's function is then decomposed into a bulk component $G_b(\bm{k}{||}, \omega)$ and a surface component $G_s(\bm{k}{||}, \omega)$, where $\bm{k}_{||}$ is the surface crystal momentum and $\omega$ is the complex frequency.

The method begins by constructing an initial principal layer,
\begin{equation}
    \begin{aligned}
    &\varepsilon_{0,mn}= \varepsilon^s_{0,mn}= \tilde{\varepsilon}^s_{0,mn} = H_{mn,11}(\bm{k}_{||})\\
    &\alpha_{0,mn} = H_{mn,12}(\bm{k}_{||})\\
    &\beta_{0,mn} = H^\dagger_{mn,12}(\bm{k}_{||})\text{,}
    \end{aligned}
\end{equation}
where the superscript $s$ denotes the surface principal layer and the tilde and superscript indicate the surface at the opposite end. The Hamiltonian matrices used in this method are the ones computed previously in Eq. \eqref{eq:surfhamij}. $\alpha_0$ and $\beta_0$ correspond to the initial inter layer hopping terms. 

The iterative algorithm explained in Ref. \cite{M-P-Lopez-Sancho_1985} follows by dividing each of the current principal layers into two principal layers, doubling the length of the chain. The new principal layers interact between them by weaker energy-dependent interactions. This equates to the following equations
\begin{equation}
    \begin{aligned}
        &\alpha_i = \alpha_{i-1}(\omega - \varepsilon_{i-1})^{-1}\alpha_{i-1}
        \\
        &\beta_i = \beta_{i-1}(\omega - \varepsilon_{i-1})^{-1}\beta_{i-1}
        \\
        &\varepsilon_i = \varepsilon_{i-1} + \alpha_{i-1}(\omega - \varepsilon_{i-1})^{-1}\beta_{i-1} + \beta_{i-1}(\omega - \varepsilon_{i-1})^{-1}\alpha_{i-1}
        \\
        &\varepsilon^s_i = \varepsilon^s_{i-1} + \alpha_{i-1}(\omega - \varepsilon_{i-1})^{-1}\beta_{i-1}
        \\
        &\tilde{\varepsilon}^s_i = \tilde{\varepsilon}^s_{i-1} + \beta_{i-1}(\omega - \varepsilon_{i-1})^{-1}\alpha_{i-1},
    \end{aligned}
\end{equation}
where, for clarity, we have omitted the $m$ and $n$ indices in each matrix. The iterative procedure is stopped when $|\varepsilon_i -\varepsilon_{i-1}| < \delta$ for both bulk and surface, where $\delta$ is small enough for the precision desired.
Once convergence has been reached, the Green's function can be approximated by
\begin{equation}
\begin{aligned}
G_s(\bm{k}_{||},\omega) \simeq (\omega-\varepsilon^s_l)^{-1} \\
\tilde{G}_s(\bm{k}_{||},\omega) \simeq (\omega-\tilde{\varepsilon}^s_l)^{-1} \\
G_b(\bm{k}_{||},\omega) \simeq (\omega-\varepsilon_l)^{-1} \text{.}
\end{aligned}
\end{equation}
where we indicate the last iteration by the index $l$.

The spectral function can be computed from the imaginary part of the trace of the converged Green's function as
\begin{equation}    A_b(\bm{k}_{||},w) = -\frac{1}{\pi}\lim_{\eta \rightarrow 0^+} \operatorname{Im}\operatorname{Tr}G_b(\bm{k}_{||},w+i\eta)\text{,}
\end{equation}
where $w$ represents the real frequency and $\eta$ a small positive number.

\section{Usage}
\label{usage}

In this section, we explain how to install \texttt{Simphony} and run the program.

\subsection{Download and compilation}

The complete \texttt{Simphony} code can be downloaded from its GitHub repository at \url{https://github.com/fballestermacia/simphony}. To compile the code, LAPACK and BLAS libraries, along with a Fortran 90 compiler, are required. For parallel execution, an MPI-enabled Fortran 90 compiler is also necessary.

After downloading, the program will be provided in a compressed folder. Once uncompressed, the code can be compiled using the standard \texttt{Makefile} located in the \texttt{simphony/src} directory, which is compatible with most architectures.Additional template \texttt{Makefile}s are provided for specific architectures, and users may modify them as needed. To compile the code, run the following command from the \texttt{simphony/src} directory:
\begin{verbatim}
    make
\end{verbatim}
will create an executable \texttt{pn.x} file inside the \texttt{simphony/bin} directory. 

\subsection{Running \texttt{Simphony}}

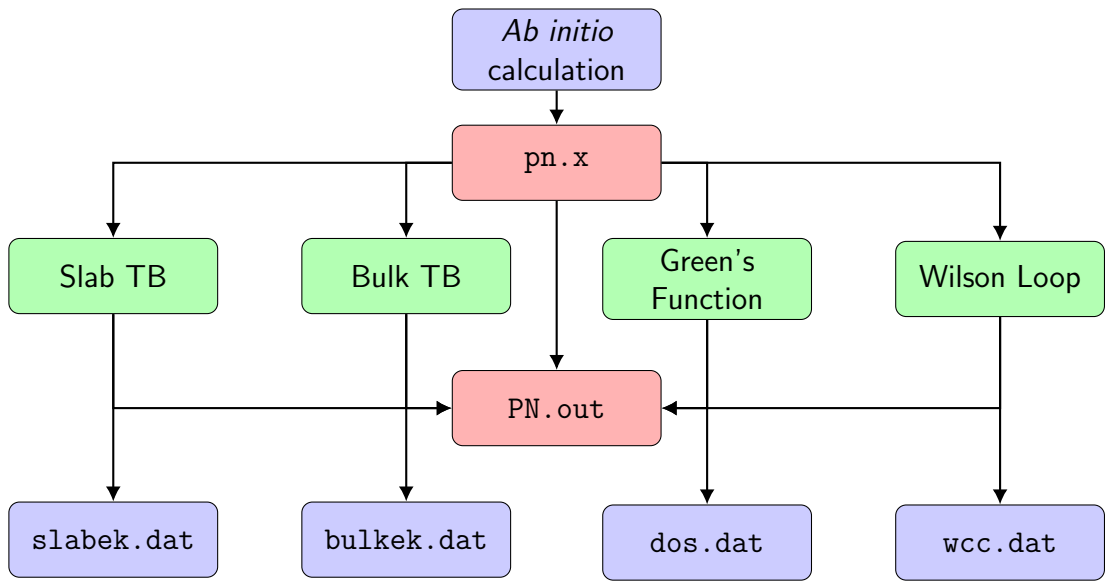
\begin{figure}[h]
    \centering
    \tikzset{%
  >={Latex[width=2mm,length=2mm]},
            base/.style = {rectangle, rounded corners, draw=black,
                           text width=2.5cm, minimum height=1cm,
                           text centered,align=center, font=\sffamily},
            blue/.style = {base,shape=rectangle,fill=blue!20},
            red/.style = {base, fill=red!30},
            green/.style = {base, fill=green!30},
            finish/.style = {base,shape=rectangle,shape border rotate=180,fill=blue!20}}

\makebox[\textwidth][c]{\begin{tikzpicture}[node distance=1cm,
    every node/.style={fill=white, font=\sffamily, minimum width=2cm}, align=center]
  \node (dft)             [blue]              {\textit{Ab initio} calculation};

  \node (runBlock)      [red, below of=dft, anchor=north] {\texttt{pn.x}};
  
  \node (bulkBlock) [green, below of=runBlock,xshift=-2cm,anchor=north] {Bulk TB};

  \node (slabBlock) [green, left of=bulkBlock, xshift=-1.5cm, anchor=east] {Slab TB};

  \node (gfBlock) [green, below of=runBlock,xshift=+2cm,anchor=north] {Green's Function};
  
  \node (wlBlock) [green, right of=gfBlock, xshift=1.5cm, anchor=west] {Wilson Loop};

  \node (outBlock) [red, below of=bulkBlock, xshift=2cm, yshift=-0.25cm, anchor=north] {\texttt{PN.out}};

  \node (slabekBlock) [blue, below of=slabBlock, yshift=-2cm, anchor=north] {\texttt{slabek.dat}};

  \node (bulkekBlock) [blue, below of=bulkBlock, yshift=-2cm, anchor=north] {\texttt{bulkek.dat}};

  \node (dosBlock) [blue, below of=gfBlock, yshift=-2cm, anchor=north] {\texttt{dos.dat}};

  \node (wccBlock) [blue, below of=wlBlock, yshift=-2cm, anchor=north] {\texttt{wcc.dat}};

  \draw[thick,->]             (dft) -- (runBlock);
  \draw[thick,->]             (runBlock) -- (outBlock);

  \draw[thick,->]             (runBlock) -| (slabBlock);
  \draw[thick,->]             (runBlock) -| (bulkBlock);
  \draw[thick,->]             (runBlock) -| (gfBlock);
  \draw[thick,->]             (runBlock) -| (wlBlock);

  \draw[thick,->]             (slabBlock) |- (outBlock);
  \draw[thick,->]             (bulkBlock) |- (outBlock);
  \draw[thick,->]             (gfBlock) |- (outBlock);
  \draw[thick,->]             (wlBlock) |- (outBlock);

  \draw[thick,->]             (slabBlock) -- (slabekBlock);
  \draw[thick,->]             (bulkBlock) -- (bulkekBlock);
  \draw[thick,->]             (gfBlock) -- (dosBlock);
  \draw[thick,->]             (wlBlock) -- (wccBlock);

  \end{tikzpicture}}
    \caption{Example workflow scheme for a typical \texttt{Simphony} calculation with the main subroutines and output files.}
    \label{fig:workflow}
\end{figure}

Running \texttt{Simphony} requires only two input files: a \texttt{wannier90\_hr.dat} file, which contains all the information about the dynamical matrix needed to build the tight-binding model (following the format described in \textsc{Wannier90}~\cite{Wannier90}), and a \texttt{pn.in} file, which provides system details, instructions, and simulation parameters.

The \texttt{wannier90\_hr.dat} file contains the hopping parameters of the TB model in real space, with the long-range dipole-dipole interaction subtracted, as in Eq. \eqref{eq:hamdynmatSR}. An executable Python script called \texttt{QE2TBDAT} can be found inside the \texttt{simphony/utility\_scripts} directory to read the output of a QuantumESPRESSO \texttt{ph.x} run and automatically generate the corresponding \texttt{wannier90\_hr.dat} file using SSCHA codes CellConstructor \cite{Monacelli2021} package.

The \texttt{pn.in} file is structured in namelists and cards. The main namelists of \texttt{Simphony} are:
\begin{itemize}
    \item \texttt{\&TB\_FILE}: Indicates the name of the \texttt{wannier90\_hr.dat} file.
    \item \texttt{\&CONTROL}: Lets \texttt{Simphony} know which subroutines to run.
    \item \texttt{\&SYSTEM}: Includes parameters when building the system, such as the number of layers in the the slab or the number of bands to include in the WCC calculation.
    \item \texttt{\&PARAMETERS}: Parameters of the simulation, such as the number of points in the grid in reciprocal space or energy range on which to compute the spectral function.
\end{itemize}

The complete list of available routines included in \texttt{Simphony} is:
\begin{itemize}
    \item \texttt{LOTO\_correction}: whether to include LO--TO splitting.
    
    \item \texttt{BulkBand\_calc}: calculate the bulk band structure along a high-symmetry path.
    \item \texttt{BulkBand\_plane\_calc}: calculate the bulk band structure in a plane.
    \item \texttt{BulkGap\_cube\_calc}: calculate the energy gap between two bands on a 3D grid of $q$-points.
    \item \texttt{BulkGap\_plane\_calc}: calculate the energy gap between two bands in a plane.
    
    \item \texttt{SlabBand\_calc}: construct a slab Hamiltonian and compute its band structure along a path.
    \item \texttt{SlabSS\_calc}: compute the surface spectral function along a path within an energy window.
    \item \texttt{SlabArc\_calc}: compute the surface spectral function on a constant-energy isosurface.
    
    \item \texttt{WireBand\_calc}: construct a ribbon Hamiltonian and compute the band structure from $-\pi$ to $\pi$.
    
    \item \texttt{Dos\_calc}: compute the density of states (DOS) within an energy window.
    
    \item \texttt{FindNodes\_calc}: locate gapless points between two bands.
    \item \texttt{BerryPhase\_calc}: compute the Berry phase along a closed loop in the 3D Brillouin zone.
    \item \texttt{BerryCurvature\_calc}: compute the Berry curvature in a 2D plane.
    \item \texttt{Chern\_3D\_calc}: compute the Wannier charge centers (WCCs) along the planes $k_1 = 0$, $k_1 = 0.5$, $k_2 = 0$, $k_2 = 0.5$, $k_3 = 0$, and $k_3 = 0.5$.
    \item \texttt{Wanniercenter\_calc}: compute WCCs in a specified plane.
    \item \texttt{WeylChirality\_calc}: compute WCCs on a sphere centered at a given point to determine Weyl node chirality.
\end{itemize}

Another Python script, called \texttt{createSimphonyInput} and included in \texttt{Simphony}, automatically generates the \texttt{pn.in} file from the output of a QuantumESPRESSO \texttt{ph.x} run. The cards in the \texttt{pn.in} file are routine-dependent, and general default values for these cards are generated by the \texttt{createSimphonyInput} script. Some cards are required for most subroutines, such as \texttt{LATTICE}, \texttt{ATOMS}, and \texttt{KPATH\_BULK}.

The \texttt{LATTICE} card is structured as follows:
\begin{verbatim}
    LATTICE
    Angstrom or Bohr
    a1    
    a2
    a3
\end{verbatim}
The \texttt{ATOMS} card has the following format:
\begin{verbatim}
    ATOMS
    Number of Atoms
    Direct or Cartesian
    Symbol     mass (in a.u.)              position
    Symbol     mass (in a.u.)              position
    ...
\end{verbatim}
And the \texttt{KPATH\_BULK} card is:
\begin{verbatim}
    KPATH_BULK
    Number of segments
    Start_label1  coordinates (direct)    End_label1  coordinates (direct)
    Start_label2  coordinates (direct)    End_label2  coordinates (direct)
    Start_label3  coordinates (direct)    End_label3  coordinates (direct)
    ...
\end{verbatim}

Examples of various input files can be found in the \texttt{simphony/examples} directory.

Then, inside a folder containing both input files, the \texttt{Simphony} executable can be run as follows:
\begin{verbatim}
    pn.x
\end{verbatim}
or in multiprocessor mode
\begin{verbatim}
    mpirun -np 4 pn.x 
\end{verbatim}

Afterward, \texttt{Simphony} will generate several output files in the current directory. The main output file, \texttt{PN.out}, contains information about the run and how \texttt{Simphony} interpreted the input files. Additionally, depending on the calculations performed, other files will be produced, primarily \texttt{.dat} and \texttt{.gnu} files. The former contains the numerical data resulting from the calculations, conveniently structured for plotting, while the latter consists of scripts used by \texttt{gnuplot}~\cite{Gnuplot} to quickly visualize the \texttt{.dat} files. A schematic flowchart detailing the main routines and output files is shown in Fig.~\ref{fig:workflow}.

\section{Examples}
\label{examples}

To illustrate the usage of \texttt{Simphony}, we present a simple Buckled Honeycomb Lattice model along with two real materials, using example input files for the calculation of their topological properties. All necessary files can be found in the \texttt{simphony/examples} directory.

\subsection{Buckled Honeycomb Lattice}

\begin{figure}
    \centering
    \includegraphics[width=\linewidth]{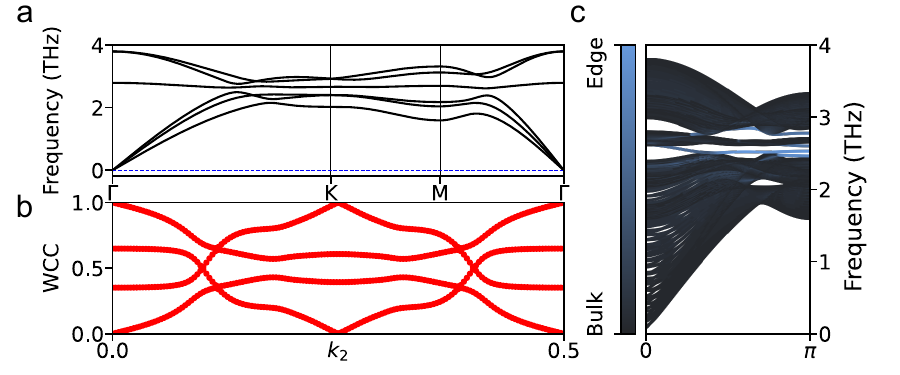}
    \caption{Example calculations for the BHL model. (a) Phonon dispersion computed by \texttt{Simphony}. (b) WCC up to the 4th band, indicating a winding number of 1. (c) Phonon dispersion of an infinite ribbon of 40 unit cells of with, color represents the projection onto the edges of the ribbon.}
    \label{fig:bhlexample}
\end{figure}

The buckled honeycomb lattice (BHL) model can present non-trivial sets of bands when considering third-nearest neighbor couplings. To showcase the usage of \texttt{Simphony} we construct a simple BHL model as in Ref. \cite{martin-paper}, up to third nearest-neighbor couplings. According to topological quantum chemistry this system is trivial, but numerical calculations show a non-zero winding number in the Wilson loop, which indicates a topological phase. 

We compute the bulk dispersion, Wannier charge center and ribbon dispersion in one single run of \texttt{Simphony}. To do so, we set the \texttt{\&CONTROL} namelist with the following variables set to true:

\begin{verbatim}
&CONTROL
BulkBand_calc         = T   
Wanniercenter_calc    = T
WireBand_calc         = T
/
\end{verbatim}

This tells \texttt{Simphony} which calculations to perform, but the program still requires the specific parameters for the simulation. An example of such an input file can be found in the \texttt{simphony/examples/BHL} directory. The following calculation-specific parameters must also be included in the input file:

\begin{verbatim}
&SYSTEM 
NSLAB1 = 40
NSLAB2 = 1
NumOccupied = 4      
/

&PARAMETERS 
Nk1 = 150            
Nk2 = 150           
/

KPLANE_BULK
0.00 0.00 0.00   ! Original point for 3D k plane
1.00 0.00 0.00  ! The first vector to define 3d k space plane
0.00 1.00 0.00  ! The second vector to define 3d k space plane

MILLER_INDEX        
1 0 0
\end{verbatim}

For the calculation of the bulk band structure, the only required parameter is the number of points per segment, specified by the \texttt{Nk1} variable. The Wannier center subroutine requires a grid of points in a plane of reciprocal space, defined by \texttt{Nk1} and \texttt{Nk2}. The \texttt{NumOccupied} variable specifies the number of occupied bands to be considered, from the first up to the \texttt{NumOccupied}-th band, for the WCC calculation. This subroutine also needs a plane on which to define the grid, provided via the \texttt{KPLANE\_BULK} card.

For the ribbon band structure, \texttt{Nk1} defines the number of points along the 1D Brillouin zone, and a surface must be specified using the \texttt{MILLER\_INDEX} card. To construct the ribbon geometry, the unit cell is repeated \texttt{NSLAB1} times along the non-periodic direction and \texttt{NSLAB2} times along the periodic direction.

In case some parameters are missing from the input file, \texttt{Simphony} uses some default values that might not be appropriate for all calculations. All details regarding the parameters used are found in the \texttt{PN.out} output file.

The results of this example are presented in Fig.\ref{fig:bhlexample}. The bulk phonon band structure, shown in Fig.\ref{fig:bhlexample}a, displays three distinct sets of bands. By performing the Wannier charge center calculation filling up to the isolated band, we obtain Fig.\ref{fig:bhlexample}b, which reveals a winding number of 1. This topological invariant indicates the presence of protected edge states, which are clearly visible in the ribbon band structure shown in Fig.\ref{fig:bhlexample}c.

\subsection{Obstructed atomic limit in AgP$_2$}
$\text{AgP}_2$ is an insulator belonging to space group  $P2_1/c$ (No. 14), thus, in the \texttt{\&CONTROL} namelist we must include:
\begin{verbatim}
LOTO_correction       = T
\end{verbatim}
for all calculations.

For this example, we will calculate the bulk bands of the system, the WCC of an obstructed atomic band representation (OABR), and the band structure of the slab. Thus, our full \texttt{\&CONTROL} namelist will be:
\begin{verbatim}
&CONTROL
LOTO_correction       = T
BulkBand_calc         = T
Wanniercenter_calc    = T
SlabBand_calc         = T
/
\end{verbatim}
All calculations can be done on the same run or separately.

\begin{figure}
    \centering
    \includegraphics[width=\linewidth]{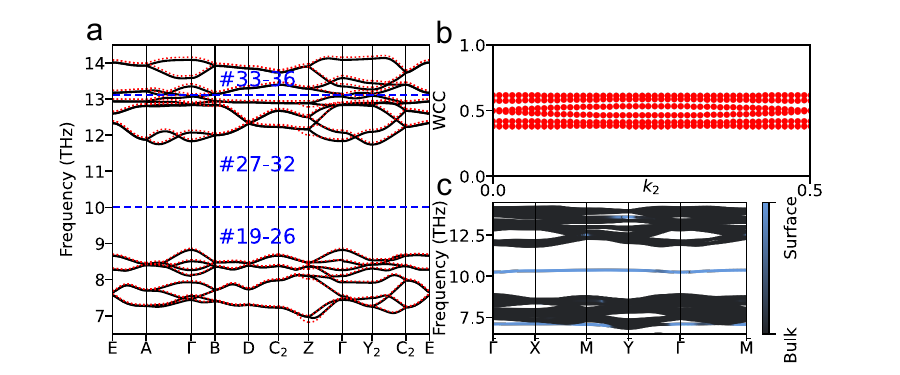}
    \caption{Example calculations for AgP$_2$. (a) Phonon band structure comparison between QuantumESPRESSO (solid black line) and \texttt{Simphony} (dashed red line). Only the top three sets of bands are shown, separated by a dashed blue line and with the included bands numerated. (b) WCC on bands 27th to 32nd on the $k_z=0$ plane. (c) Slab band dispersion on the (001) surface.}
    \label{fig:agp2example}
\end{figure}

For the \texttt{BulkBand\_calc} and the \texttt{SlabBand\_calc } tags we need to set the Nk1 tag in the \texttt{\&PARAMETERS} namelist. The resulting calculation for the bulk band structure can be seen in Fig. \ref{fig:agp2example}a, showing clear agreement with \textit{ab initio} results. These were performed using QuantumESPRESSO, in a $4\times4\times4$ grid of $q$-points, using a PBE functional and a $8\times8\times8$ grid of $k$-points for the SCF calculation. The \texttt{wannier90\_hr.dat} file was created using the \texttt{QE2TBDAT} python script included in \texttt{Simphony}, which automatically removes the long-range dipole-dipole interaction from the dynamical matrix in real space. An agreement between the \textit{ab initio} band structure and the \texttt{Simphony} band structure indicates an equivalent implementation of the LO-TO splitting.

We also need \texttt{Nk2} for the WCC calculation, so we must include both in the \texttt{\&PARAMETERS} namelist
\begin{verbatim}
&PARAMETERS    
Nk1 = 50           
Nk2 = 50                   
/
\end{verbatim}

For the \texttt{\&SYSTEM} namelist, the width of the slab to be computed must be set using the \texttt{NSLAB} parameter. For the WCC calculation, the \texttt{NumOccupied} parameter should be set to the index of the highest band to be considered occupied. Thus:
\begin{verbatim}
&SYSTEM 
NSLAB = 10
NumOccupied = 32
/
\end{verbatim}
However, this will consider the subspace spanned from the first band to the 32nd band, and since this subspace includes several independent sets of bands, it is often more convenient to consider isolated sets of bands for the WCC calculation. In this case, the relevant set corresponds to bands 27 through 32, which can be selected using the \texttt{SELECTED\_OCCUPIED\_BANDS} card as follows:

\begin{verbatim}
SELECTED_OCCUPIED_BANDS
27-32
\end{verbatim}
According to the results in Fig.~\ref{fig:agp2example}b, this set of bands corresponds to an OABR, which, due to the bulk–edge correspondence, leads to a surface state. Thus, we select the (001) surface for our slab calculations by including the following card:

\begin{verbatim}
SURFACE
1 0 0
0 1 0
\end{verbatim}
which selects the surface defined by vectors $\bm{k}_1$ and $\bm{k}_2$. Then, the \texttt{SlabBand\_calc } leads to Fig. \ref{fig:agp2example}c, which shows a clear surface state between the two sets of bands. For clarity we only show the projection onto the surface, as in Eq. \eqref{eq:projectionSS}, but \texttt{Simphony} is capable of projecting onto any of the two surfaces of the slab.

\subsection{Weyl nodes in Al$_2$ZnTe$_4$}
\begin{figure}
    \centering
    \includegraphics[width=\linewidth]{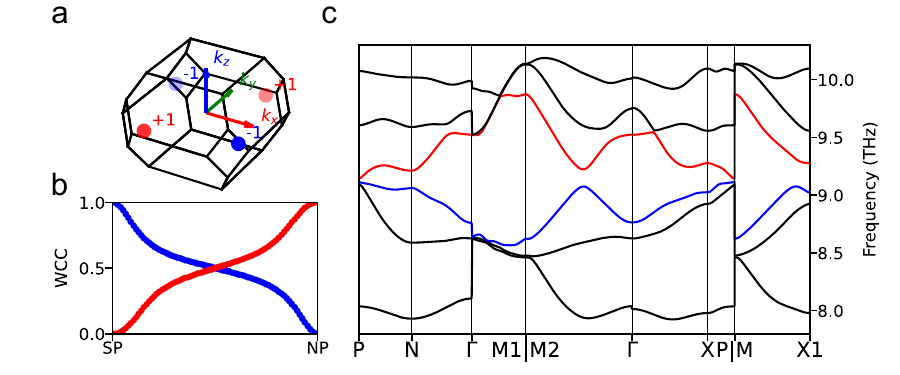}
    \caption{Example calculations for Al$_2$ZnTe$_4$. (a) First Brillouin zone and the four Weyl nodes located in the $k_z = 0$ plane between the 18th and 19th bands. The associated chirality is indicated next to each node. Red (blue) nodes correspond to positive (negative) chirality and act as sources (sinks) of Berry curvature. (b) Weyl chirality calculated on a sphere surrounding each node. Colors correspond to the nodes shown in (a), and the winding number reflects the chirality of each node. (c) Phonon band dispersion in a selected energy window of Al$_2$ZnTe$_4$ as computed by \texttt{Simphony}. The 18th (19th) band is highlighted in blue (red).
 }
    \label{fig:al2znte4example}
\end{figure}

In the final example, we study the Weyl nodes between the 18th and 19th bands of Al$_2$ZnTe$_4$, a compound belonging to space group $I\bar{4}$ (No.~82). As this compound is an insulator, the LO--TO correction must be taken into account. The phonon bands are shown in Fig.~\ref{fig:al2znte4example}c. Since the system is not cubic, discontinuities appear at $\Gamma$ due to LO--TO splitting.

We begin by showcasing the \texttt{FindNodes\_calc} subroutine, which locates gapless points using Nelder and Mead's downhill simplex method~\cite{nelder-1965}. To do so, we need to define a sufficiently dense $q$-point mesh—typically, 15 points in each direction are sufficient—by setting the \texttt{Nk1}, \texttt{Nk2}, and \texttt{Nk3} parameters accordingly. Additionally, a threshold for identifying gapless points can be specified using the \texttt{Gap\_threshold} tag in the \texttt{\&PARAMETERS} namelist. For example:

\begin{verbatim}
&CONTROL
LOTO_correction       = T 
FindNodes_calc        = T
/

&SYSTEM
NumOccupied = 18 
/

&PARAMETERS
Nk1 = 15   
Nk2 = 15 
Nk3 = 15  
Gap_threshold = 0.000001   !in THz
/
\end{verbatim}

For simplicity, as done in Ref. \cite{topophodatabase}, we will consider only the Weyl nodes located at the $k_z=0$ plane. The \texttt{FindNodes\_calc} routine writes the location of the gapless points in a file named \texttt{Nodes.dat}, the location of these four points is shown in Fig. \ref{fig:al2znte4example}a.

Then, we take the coordinates of these points and include them in the \texttt{WEYL\_CHIRALITY} card as:
\begin{verbatim}
WEYL_CHIRALITY
4          
Direct     
0.00001       
 0.40836109   -0.40491678   -0.09378520
-0.40836104    0.40491685    0.09378513
-0.11256878    0.06227720   -0.38968016
 0.11256874   -0.06227727    0.38968020
\end{verbatim}
The first line indicates the number of points to be computed, the second line states whether the coordinates are in direct or Cartesian coordinates, the third line corresponds with the radius of the sphere surrounding each node on which we will perform the WCC. The following lines are the coordinates of each point. By setting \texttt{WeylChirality\_calc    = T} in the \texttt{\&CONTROL} namelist we can obtain the winding number of each Weyl node, as shown in Fig. \ref{fig:al2znte4example}c. 

\begin{figure}
    \centering
    \includegraphics[width=\linewidth]{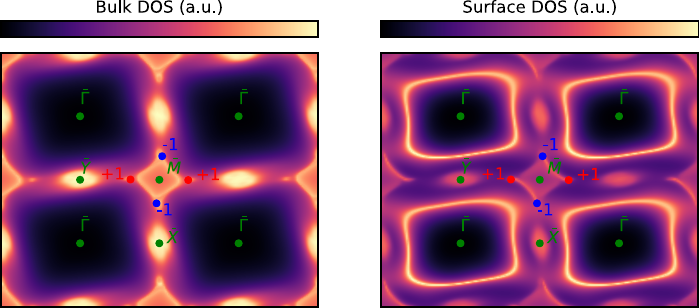}
    \caption{Energy contour at 9.21~THz of Al$_2$ZnTe$_4$ on the (001) surface, obtained using the bulk (left) and surface (right) Green's functions. A $2\times 2$ surface Brillouin zone is shown, with the two-dimensional high-symmetry points indicated. The four Weyl nodes are marked with colors and chiralities corresponding to Fig.~\ref{fig:al2znte4example}. In the center of the right panel, phonon surface arcs connect the Weyl nodes.}
    \label{fig:al2znte4example2}
\end{figure}

To obtain the energy contour shown in Fig. \ref{fig:al2znte4example2} we must indicate the grid of points on which to perform the calculation, the energy and broadening for the contour and the surface on which to do so. The input file should be similar to
\begin{verbatim}
&CONTROL
LOTO_correction       = T 
SlabArc_calc          = T
/



&PARAMETERS
Nk1 = 100   
Nk2 = 100 
Eta_Arc = 0.003   
E_arc = 9.210 
/
\end{verbatim}

The results of this subroutine are shown in Fig. \ref{fig:al2znte4example2}. Surface states connecting the Weyl nodes in the (001) surface Brillouin zone are visible, in accordance with the WCC calculation.

\section{Conclusion}
We present \texttt{Simphony}, a new open-source software based on maximally localized lattice Wannier functions, which extends the Wannier tight-binding framework commonly used for electrons to the case of lattice vibrations. Particular emphasis is placed on the treatment of long-range interactions that arise in polar materials. We demonstrate the capabilities of the software through three examples: the topological diagnosis of a buckled honeycomb lattice toy model, the surface states arising from an OABR in AgP$_2$, and the Weyl nodes and their associated chirality in Al$_2$ZnTe$_4$.

\section*{Acknowledgments}
M.G.V received financial support from the Canada Excellence Research Chairs Program for Topological Quantum Matter, NSERC Quantum Alliance France-Canada and Diputación Foral de Gipuzkoa Programa Mujeres y Ciencia.  M.G.V and F.B. thank support PID2022-142008NB-I00 projects funded by MICIU/AEI/10.13039/501100011033 and FEDER, UE and the Ministry for Digital Transformation and of Civil Service of the Spanish Government through the QUANTUM ENIA project call - Quantum Spain project, and by the European Union through the Recovery, Transformation and Resilience Plan - NextGenerationEU within the framework of the Digital Spain 2026 Agenda. I.E. acknowledges funding from the Spanish Ministry of Science and Innovation (Grant No. PID2022-142861NA-I00) and the Department of Education, Universities and Research of the Basque Government and the University of the Basque Country (Grant No. IT1527-22).

\bibliographystyle{elsarticle-num}
\bibliography{references.bib}

\end{document}